\documentclass[aps,prb,twocolumn,superscriptaddress]{revtex4-2}
\usepackage{graphicx}
\usepackage{dcolumn}
\usepackage{bm}
\usepackage{gensymb} 
\usepackage{amsmath,amssymb,bm,amsfonts,mathrsfs,epsfig,amsbsy,verbatim,bbm,mathtools,slashed} 
\usepackage{color}

\begin{document}

\title{Symmetry-breaking normal state response and surface superconductivity in topological semimetal YPtBi}

\author{Hyunsoo Kim}
\affiliation{Maryland Quantum Materials Center, Department of Physics, University of Maryland, College Park, Maryland 20742, USA}
\affiliation{Department of Physics, Missouri University of Science and Technology, Rolla, Missouri 65409, USA}

\author{Tristin Metz}
\affiliation{Maryland Quantum Materials Center, Department of Physics, University of Maryland, College Park, Maryland 20742, USA}

\author{Halyna Hodovanets}
\affiliation{Maryland Quantum Materials Center, Department of Physics, University of Maryland, College Park, Maryland 20742, USA}
\affiliation{Department of Physics, Missouri University of Science and Technology, Rolla, Missouri 65409, USA}

\author{Daniel Kraft}
\affiliation{Maryland Quantum Materials Center, Department of Physics, University of Maryland, College Park, Maryland 20742, USA}

\author{Kefeng Wang}
\affiliation{Maryland Quantum Materials Center, Department of Physics, University of Maryland, College Park, Maryland 20742, USA}

\author{Yun Suk Eo}
\affiliation{Maryland Quantum Materials Center, Department of Physics, University of Maryland, College Park, Maryland 20742, USA}
\affiliation{Department of Physics \& Astronomy, Texas Tech University, Lubbock, Texas 79410, USA}

\author{Johnpierre Paglione}
\affiliation{Maryland Quantum Materials Center, Department of Physics, University of Maryland, College Park, Maryland 20742, USA}
\affiliation{Canadian Institute for Advanced Research, Toronto, Ontario M5G 1Z8, Canada}

\date{\today}

\begin{abstract}
Most of the half-Heusler RPtBi compounds (R=rare earth) host various surface states due to spin-orbit coupling driven topological band structure.
While recent ARPES measurements ubiquitously reported the existence of surface states in RPtBi, their evidence by other experimental techniques remains elusive. 
Here we report the angle-dependent magnetic field response of electrical transport properties of YPtBi in both the normal and superconducting states.
The angle dependence of both magnetoresistance and the superconducting upper critical field breaks the rotational symmetry of the cubic crystal structure, and the angle between the applied magnetic field and the measurement plane of a plate-like sample prevails. 
Furthermore, the measured upper critical field is notably higher than the bulk response for an in-plane magnetic field configuration, suggesting the presence of quasi-2D superconductivity. 
Our work suggests the transport properties cannot be explained solely by the bulk carrier response, requiring robust normal and superconducting surface states to flourish in YPtBi.
\end{abstract}



\maketitle

\section{Introduction}
It has long been known that the boundary between two solid-state systems can host fascinating electronic states that can fundamentally differ from those of the host materials \cite{Shockley1939}. 
Traditionally, such boundary states are best demonstrated at the interface of different semiconductors, and the systematic control of the semiconductor heterostructure led to the discovery of topological edge states in the Quantum Hall (QH) phase \cite{Klitzing1980}. 
Over the past decades, the experimental realization of the Quantum Spin Hall (QSH) effect \cite{Bernevig2006,Koenig2007}, topological insulator (TI) phases \cite{Hasan2010} and more recently the fractional quantum anomalous Hall effect (FQAHE) \cite{Cai2023,Park2023,Zeng2023,Xu2023,Lu2024} have launched intense research activities investigating of various surface states in compounds with finite topological invariants \cite{DKim2012,DKim2012NatPhys,DasSarma2013}. 
Such topological surface states provide a unique transport mechanism immune to back-scattering due to its time-reversal invariant nature, making it a promising candidate for various technological applications. 
Superconducting topological surface states are of particular interest because of the possibility of harboring an exotic Majorana surface fluid \cite{Timm2017,Yang2017,Ghorashi2017,Wu2020,Zhang2021}, hosting novel phenomena such as the Klein paradox \cite{Young2009,Lee2019}, and harboring intrinsic surface superconductivity mediated by electron-phonon interaction as predicted for a TI \cite{DasSarma2013}.
Due to the low-dimensional nature of a surface superconducting state, it is protected from orbital pair-breaking in the case of a parallel applied field \cite{Tinkham2004}, and can withstand a high in-plane magnetic field.
While surface nucleation of the classic superconducting ``$H_{c3}$'' phase is well known to manifest a higher critical field $H_{c3}(0) \approx 1.7 H_{c2}(0)$ \cite{Saint-James1963, Tinkham2004, Kim2013}, topological surface superconductivity with strong spin-orbit coupling can also lead to extremely high upper critical fields exceeding the Pauli paramagnetic limiting field \cite{Lu2015,Xi2016,Saito2016}. 

The half-Heusler RPtBi (R=rare earth) compounds have attracted a great deal of attention because of possible topological surface states \cite{Chadov2010,Lin2010,Feng2010}. 
The theoretical band structure is consistent with a topological semimetal driven by the strong spin-orbit coupling which inverts the $s$-orbital derived $\Gamma_6$ band and $p$-orbital derived $\Gamma_8$ band. 
The theoretical chemical potential lies on a quadratically touching point of the $\Gamma_8$ band, which makes YPtBi a zero-gap semiconductor or semimetal \cite{Chadov2010,Lin2010,Brydon2016,Kim2018}. 
On the other hand, the experimentally determined chemical potential sits about 35 meV below the touching point \cite{Butch2011,Kim2018,Kim2022}. 
The quasiparticles are best described by the $j=3/2$ Luttinger-Kohn Hamiltonian \cite{Brydon2016,Cano2017,Kim2018,Kim2022}, which was experimentally verified by recent quantum oscillations experiments \cite{Kim2022}. 
These $j=3/2$ quasiparticles can form Cooper pairs with total effective spin angular momentum $J$ beyond spin triplet  states $J=1$ \cite{Brydon2016,Savary2017,Venderbos2018}, and a $J=3$ septet pairing state was proposed to reconcile evidence for unconventional line-nodal superconductivity \cite{Brydon2016,Kim2018,Ishihara2021}. This so-called ``high-spin'' superconductivity offers the potential to realize various topological superconducting states \cite{Brydon2016,Venderbos2018}, and can harbor a Majorana surface fluid \cite{Timm2017,Yang2017,Ghorashi2017,Wu2020,Zhang2021} which can be utilized for fault-tolerant topological quantum information technology.

\begin{figure*}[!ht]
\includegraphics[width=0.99\linewidth]{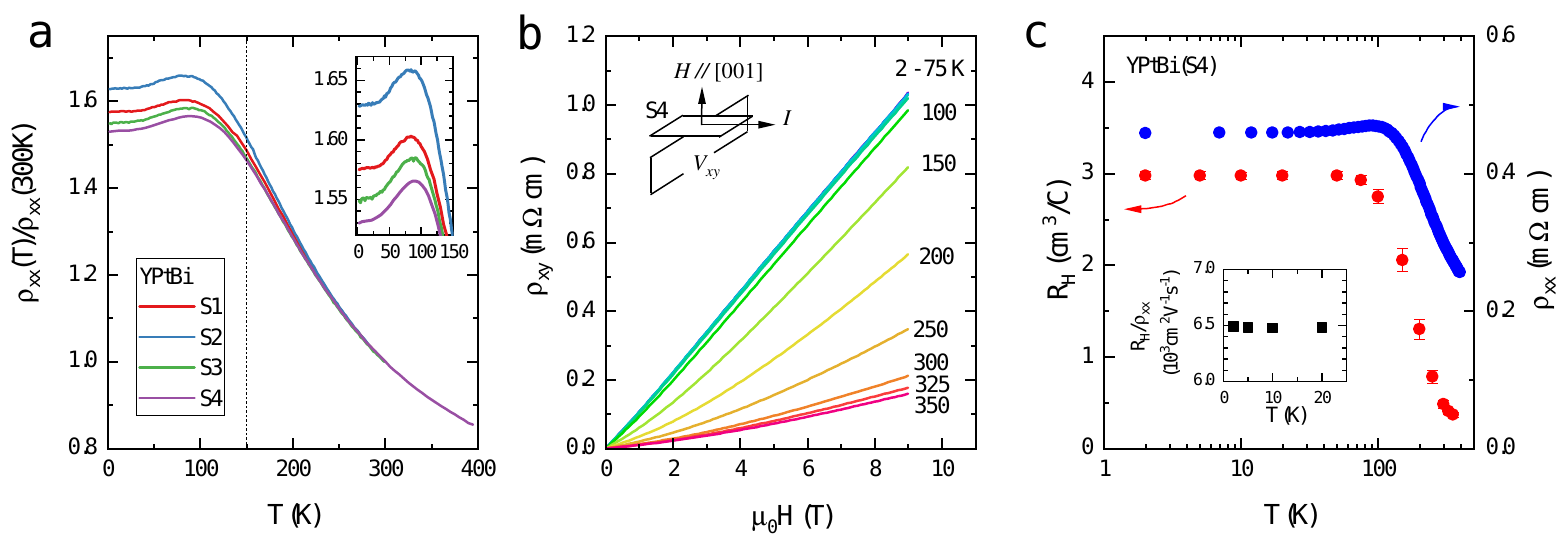}%
\caption{\label{fig1} {\bf Resistivity and Hall measurements in YPtBi. }(a) Normalized temperature-dependent resistivity $\rho_{xx}(T)/\rho_{xx}(300\textmd{K})$. Inset: low-temperature part. (b) Field-dependent Hall resistivity $\rho_{xy}(H)$ in sample S4 at various temperatures with $H\parallel$[100]. (c) Comparison of temperature-dependent $R_H(T)$ to $\rho_{xx}(T)$.}
\end{figure*}

In YPtBi, multiple ARPES experiments found metallic surface states \cite{Liu2011,Liu2016,Kim2018,Hosen2020}, but the topological nature of the surface states is not transparent. 
Furthermore, the detailed structure of the bulk $\Gamma_8$ band near the touching point needs to be experimentally verified, which depends on the choice of many-body potential energy term in first-principle calculations \cite{Feng2010}. 
Understanding the band structure and superconducting properties in YPtBi would shed light on the current unresolved issues. Moreover, 
observations of an enhanced onset of the superconducting transition temperature $T_c$ \cite{Baek2015} and indications of an unconventional surface superconducting state \cite{persky2023possible} in YPtBi suggest 
a possible topological chiral state to thrive on the surface of this material \cite{schwemmer2022chiral}.
In this work, we investigate the angle-dependent electrical transport properties in normal and superconducting states in topological semimetal YPtBi to understand the symmetry constraints on these novel phases.

\section{Experiment}
Single crystals of the half-Heusler YPtBi were synthesized by utilizing a high-temperature flux technique in molten Bi with a starting ratio of 1:1:20 for Y, Pt, and Bi, respectively. 
YPtBi single crystals were grown out of molten Bi with starting composition Y:Pt:Bi = 1:1:20 (atomic ratio) \cite{Butch2011,Kim2018}. 
The starting materials Y ingot (99.9\%), Pt powder (99.95\%), and Bi chunk (99.999\%) were put into an alumina crucible, and the crucible was sealed inside an evacuated quartz ampule. 
The ampule was heated slowly to 1150\degree C, kept for 10 hours, and then cooled down to 500\degree C at a 3\degree C/hour rate, where the excess of molten Bi was decanted by centrifugation. The MgAgAs-type half-Heusler phase ($F\bar{4}3m$) of the single crystal samples was confirmed by powder x-ray diffraction pattern. 
A single-crystal x-ray technique was utilized to determine the crystallographic directions \cite{Kim2018}.

For transport measurements, samples were mechanically polished into bar- or plate-like shapes with typical dimensions 1.0$\times$0.3$\times$0.035~mm$^3$. 
Electrical contacts to high-purity silver wire leads were attached by using silver paste.
Typical contact resistances were of order $\sim 1~\Omega$ or less.
Magneto-transport measurements were performed using a Quantum Design Physical Property Measurement System equipped with a 14~T superconducting magnet and a single-axis rotator. 
Electrical resistance and Hall voltage were determined by using an ac resistance bridge.
For measurements of the angle-dependent superconducting upper critical field, we utilized an Oxford instruments dilution refrigerator and a 5~T two-axis vector magnet.

\section{Results}

Figure \ref{fig1}(a) presents the temperature-dependent resistivity $\rho_{xx}(T)$ of four different YPtBi samples in the absence of magnetic fields. 
All $\rho_{xx}(T)$ curves are normalized to the resistivity values determined at 300 K, and exhibit a nearly identical temperature dependence, with a negative slope at high temperatures, an inflection point near 150 K, and a slight variation of the residual resistivity values $\rho_{xx}(0)$. 
All samples consistently show a broad maximum of around 80 K. 
The low-temperature part of $\rho_{xx}(T)$ below $T=150$ K is shown in the inset which clearly displays the positive slope of $\rho_{xx}(T)$ at low temperatures with much weaker temperature variation.

Figure \ref{fig1}(b) shows the field-dependent Hall resistivity $\rho_{xy}(H)$ in sample S4 at various temperatures with $H\parallel$[001]. 
For temperatures up to 75 K, $\rho_{xy}(H)$ exhibits nearly linear field variation, but a noticeable curvature in $\rho_{xy}(H)$ appears with temperatures above 100 K. 
Overall temperature dependence becomes weaker with increasing temperature.
The temperature-dependent Hall coefficient $R_H(T)=V_{xy} t/I\mu_0 H$ for sample S4 is shown in Figure \ref{fig1}(c). 
Here $V_{xy}$ is the measured voltage with applied $I$ and $H$, and $t$ is the thickness of the sample.
The measured $V_{xy}(H)$ is not linear in $H$ at higher temperatures, and $R_H$ was determined by fitting a linear line to the data up to 9 T. 
The error bars represent fitting errors. 
$R_H(T)$ strongly depends on temperature with a negative slope between 100 K and 350 K, but it is nearly constant below 100 K. 
This behavior is reminiscent of $\rho_{xx}(T)$. 
For comparison $\rho_{xx}(T)$ is shown, which was determined in the same sample, and they exhibit qualitatively the same temperature variation \cite{Butch2011}.
Therefore, the non-metallic behavior of $\rho_{xx}(T)$ is partially explained by the temperature variation of the carrier density. 
As shown in the Hall effect (see Fig. \ref{fig1}(b)), $\rho_{xy}(H)$ is consistent with a hole-like single band for temperatures below 75 K. 
The hole mobility can be estimated by using the relation, $R_H(T)/\rho_{xx}(T)$. 
The inset of Fig. \ref{fig1}(c) shows the calculated $\mu_h$ which is about $6.5\times 10^{3}$ cm$^2$V$^{-1}$s$^{-1}$ up to 20 K.

\begin{figure}[!t]
\includegraphics[width=1\linewidth]{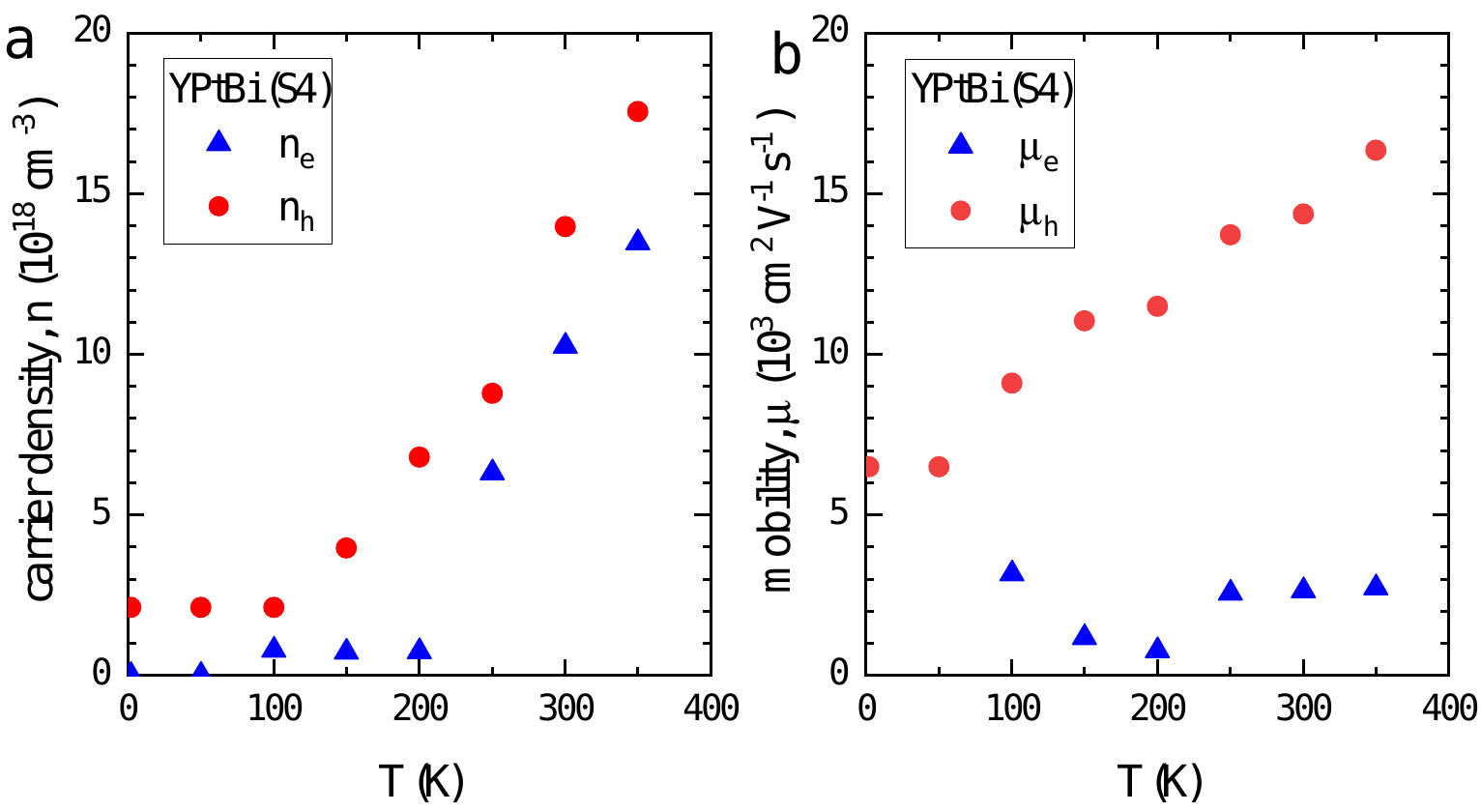}%
\caption{\label{fig2} {\bf Temperature-dependent carrier density and mobility.} (a) Carrier densities $n_h$ and $n_e$ in YPtBi at various temperatures, which are determined by fitting the two-carrier model (Eq. \ref{eq1}) to the experimental Hall data shown in Fig. \ref{fig1}. (b) Mobilities $\mu_h$ and $\mu_e$ are similarly determined at various temperatures.}
\end{figure}

In a single-band approximation, the carrier density can be determined by the relation, $R_H=1/en_h$, which is $n_h = 2.1\times 10^{18}$ cm$^{-3}$ at $T=2$ K. 
However, we cannot employ this approximation at higher temperatures above 100 K because of the pronounced curvature. The observed behavior of $\rho_{xy}(H)$ in Fig. \ref{fig1}(b) is consistent with the appearance of the electron-like carrier with increasing temperature.
Temperature dependence of the carrier density $n$ and mobility $\mu$ of electron-and hole-like carriers are determined by fitting the two-band model to the experimental $\rho_{xy}(H)$ by using the relation \cite{Luo2016}:
\begin{equation}\label{eq1}
\rho_{yx}(B)=\frac{B}{e}\frac{(n_h\mu_h^2-n_e\mu_e^2)+(n_h-n_e)\mu_h^2\mu_e^2 B^2}{(n_h\mu_h+n_e\mu_e)^2+(n_h-n_e)^2\mu_h^2\mu_e^2 B^2}
\end{equation}
The determined $n_e$ and $n_h$ are displayed in Fig.~\ref{fig2}(a). Below $T\approx 100$ K, only the hole-like carriers are active, but the electron-like carriers start populating above 100 K. 
$n_e$ is comparable to $n_h$ at high temperatures above 200 K. 
Our results indicate that YPtBi is a nearly compensated semimetal at room temperature, which explains the larger non-saturating magnetoresistance \cite{Shekhar2012,Shekhar2012prb,Shekhar2013,Shekhar2018}.

The temperature-dependent mobility of the electron-like carrier $\mu_e$ and the hole-like carrier $\mu_h$ are displayed in Fig.~\ref{fig2}(b).
Whereas $n_h$ increases monotonically with temperature, $\mu_e$ exhibits only weak variation over the entire temperature range. 

\begin{figure*}[!t]
\includegraphics[width=0.7\linewidth]{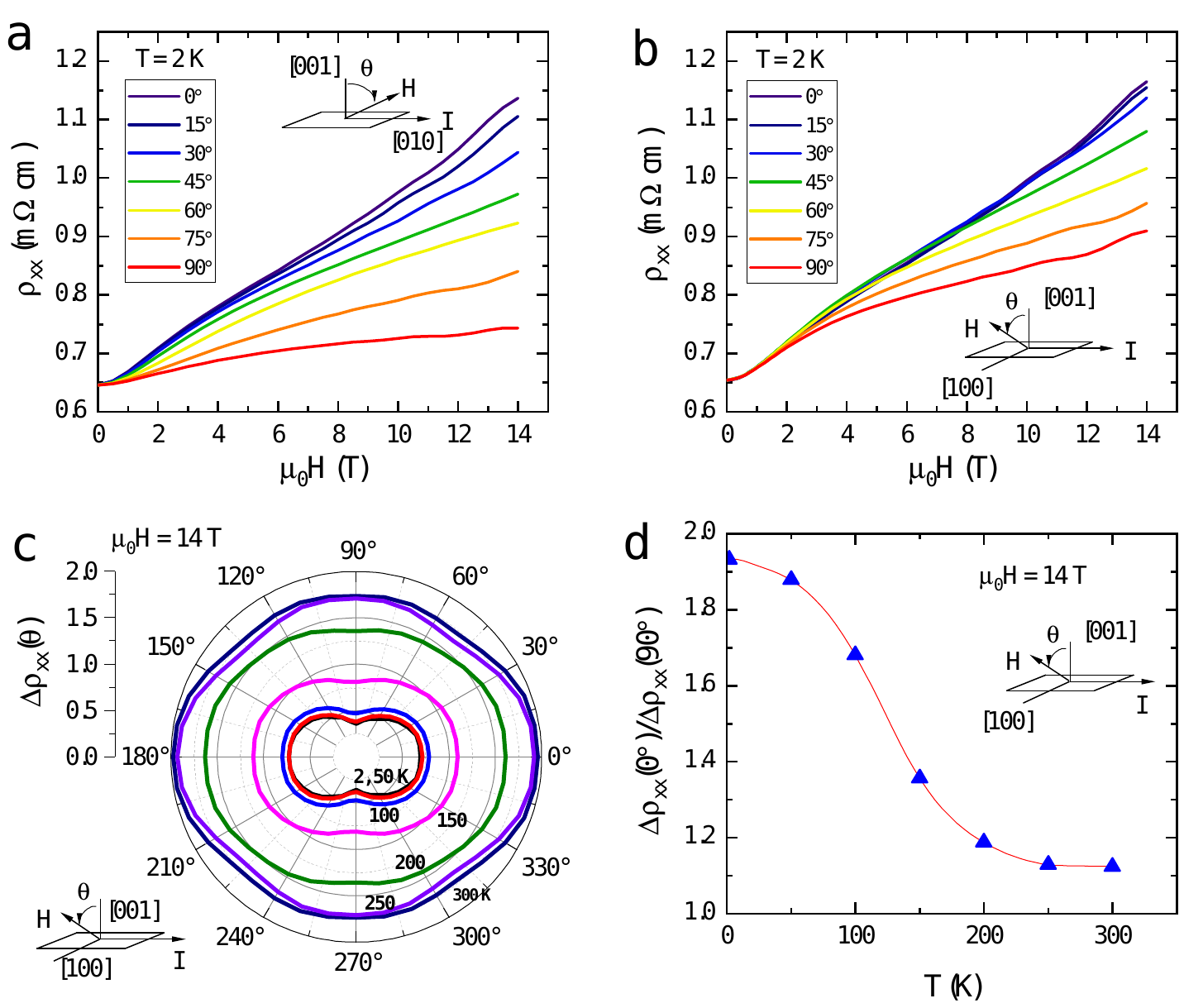}%
\caption{\label{fig3} {\bf Angle-dependent magnetoresistance in YPtBi.} (a) Field-dependent $\rho_{xx}(H)$ with various angles determined at $T=2$ K. Field was rotated from [001] to [010] as indicated in inset. (b) Similar $\rho_{xx}(H)$ data with field rotation transverse to the electric current from [001] to [100]. (c) $\Delta \rho_{xx}(\theta)=[\rho_{xx}(14 \textmd{ T})-\rho_{xx}(0 \textmd{ T})]/\rho_{xx}(0 \textmd{ T})$ as a function of the angle at various temperatures with a magnetic field rotating around the current direction. (d) Temperature variation of the magnetoresistance ratio $\Delta\rho_{xx}(0\degree)/\Delta\rho_{xx}(90\degree)$ determined at 14 T.
}
\end{figure*}

Figure \ref{fig3}(a) shows field-dependent $\rho_{xx}(H)$ in S1 at $T=2$ K with various angles. 
The sample is cut out of the (001) plane. 
The rotation of $H$ is indicated in the inset. 
The angle represents a polar angle measured from [001], normal to the sample plane, to [010] direction which is the electric current direction. 
In this configuration, the angle-dependent $\rho_{xx}$ is dominated by the Lorentz force, and magnetoresistance (MR) is zero in principle when $\theta=90\degree$. 
In our experiment with $\theta=90\degree$, $\rho_{xx}$ moderately increases with $H$, yielding $\rho_{xx}(14\textmd{ T})\approx 1.14\rho_{xx}(0\textmd{ T})$. 
We define $\Delta\rho_{xx}(\theta)=[\rho_{xx}(14\textmd{ T})-\rho_{xx}(0\textmd{ T})]/\rho_{xx}(0\textmd{ T})$ at a given angle $\theta$. At $T=2$ K, $\Delta\rho_{xx}(0\degree)/\Delta\rho_{xx}(90\degree)\approx 5.6$, which can be accounted for by the angle-dependent Lorentz force. 
The Shubnikov-de Haas (SdH) effect is visible in the high field region, and the angle-dependent SdH effect is consistent with the presence of a $j=3/2$ quasiparticle Fermi surface in YPtBi \cite{Kim2022}.

Figure \ref{fig3}(b) shows the field-variation of $\rho_{xx}(H)$ in the same sample S1 at $T=2$ K with a rotating field from [001] to [100] as depicted in the inset. 
In this experiment, however, the orientation of $H$ remains perpendicular to the direction of the electric current, and the nominal contribution of the Lorentz force remains uniform upon rotation. 
Since the angular variation of intrinsic $\rho_{xx}$ will be governed by the microscopic parameters such as effective mass $m^*$, Fermi velocity $v_F$, etc, the rotational symmetry in $\rho_{xx}$ should follow the crystallographic symmetry. 
In this case, it would exhibit a four-fold symmetry with $\rho_{xx}^{\textmd{\tiny [001]}}= \rho_{xx}^{\textmd{\tiny [100]}}$. However, we observed $\rho_{xx}^{\textmd{\tiny [100]}} (H) = 1.27 \rho_{xx}^{\textmd{\tiny [100]}}(H)$ with $\mu_0H=14$ T, and therefore $\Delta \rho_{xx}(\theta)$ breaks rotational symmetry.

We investigated the anomalous rotational symmetry of MR at various temperatures. 
Figure \ref{fig3}(c) is a polar plot of the angle-dependent $\Delta\rho_{xx}(\theta)$ at various temperatures with a magnetic field rotating around the current direction. 
The angular variation exhibits a two-fold symmetry at all measured temperatures, but it is much more pronounced at low temperatures. 
Between 2 K and 150 K, $\Delta\rho_{xx}(\theta)$ exhibits a butterfly-like shape, most notably below 50 K where the anisotropy remains nearly temperature-independent to the lowest temperatures. 
At high temperatures, $\Delta\rho_{xx}(\theta)$ exhibits a nearly four-fold symmetry and is notably smaller when a field is applied along the [110] equivalent orientations.
The temperature variation of the ratio $\Delta\rho_{xx}(0\degree)/\Delta\rho_{xx}(90\degree)$, which reflects the two-fold symmetry breaking anisotropy, is shown in Figure \ref{fig3}(d). Starting close to unity at room temperature, the anisotropy rapidly increases upon cooling, reaching nearly a factor of two at low temperatures. 
This observation suggests that additional factors need to be considered to explain the measured angular variation of magnetoresistance. 

\begin{figure*}[!ht]
\includegraphics[width=0.9\linewidth]{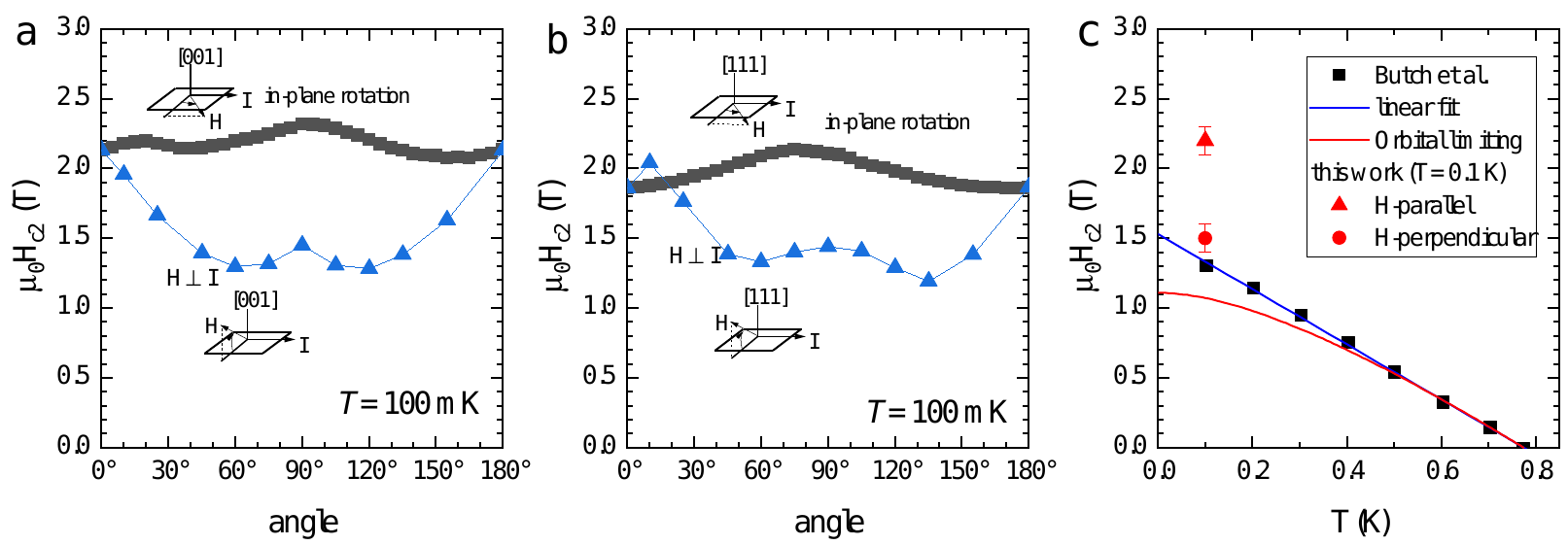}%
\caption{\label{fig4} {\bf Anisotropy of superconducting upper critical field $H_{c2}$ at $T=100$ mK in YPtBi.} The symbols represent $\mu_0 H$ determined with a magnetic field rotation as depicted in the inset for (a) the (001)-sample and (b) the (111)-sample.
}
\end{figure*}

To investigate the effect of the anomalous rotational symmetry on the superconducting state, we determined the angle-dependence of the superconducting upper critical field $H_{c2}$ by electrical transport measurements. 
Figure \ref{fig4} shows the angular variation of $H_{c2}(T)$ in the samples with (001) and (111) planes. 
At each angle, $H_{c2}$ was determined at the 50\% of superconducting-to-normal state transition while ramping up $H$ at a fixed temperature of $T=100$ mK.
Figure \ref{fig4}(a) displays the results in the (001)-sample. 
The black square symbols represent $H_{c2}$ with a magnetic field rotating in the sample plane as indicated in the inset. 
The angle-dependent $H_{c2}$ exhibits a nearly four-fold symmetry with a weak variation around $\mu_0H_{c2}(0)=2.2$ T that is significantly higher than the reported 1.5 T \cite{Butch2011}. 
The blue triangle symbols represent $H_{c2}$ with a magnetic field rotating around the electric current, starting from the direction parallel to the sample plane as indicated in the inset. 
With increasing angle, $H_{c2}$ rapidly decreases. The angle-dependence clearly exhibits a two-fold symmetry with about 30\% reduction in $H_{c2}$ when the magnetic field is applied perpendicular to the sample plane.

Figure \ref{fig4}(b) shows the results from similar experiments with a sample prepared from the (111) plane. 
The black square symbols represent $H_{c2}$ from the in-plane rotation experiments while the blue triangle symbols represent that from the rotation around the electrical current. 
The rotation is depicted in the inset for both rotation experiments. 
$H_{c2}$ in the experiment with in-plane rotation shows overall higher values similar to the results from the (001)-sample. 
In both (001)-and (111)-samples, angle-dependent $H_{c2}$ is dominated by the angle between the applied $H$ and the sample plane, irrespective to the crystallographic direction similar to the angle-dependence of magnetoresistance in the normal state.

\section{discussion}

The observed breaking of the rotational symmetry of cubic crystal in both normal and superconducting states by our transport measurements indicates the presence of a 2D state.
Several recent ARPES studies consistently showed a large chemical potential shift of about 300 meV at the surface \cite{Liu2011,Liu2016,Kim2018,Hosen2020} which is in stark contrast to the bulk chemical potential determined by Hall effect and quantum oscillation experiments \cite{Butch2011,Kim2018,Kim2022}. 
This leads to the possibility of different electronic phases near the surface and points out that the measured resistivity might not be solely from the bulk. 
The longitudinal transport is likely carried by both bulk and surface, i.e., two-channel parallel transport, and the contribution of a surface component is rather significant at low temperatures (see Fig. 3).
The local maximum around 80 K is probably associated with the crossover of a dominant component from non-metallic bulk at high temperatures to metallic surface at low temperatures, similarly with the onset of a low-temperature plateau in the topological Kondo insulator due to the surface state \cite{Wolgast2013,Syers2015,Eo2019}.
However, the significant charge imbalance at the surface due to band bending in YPtBi \cite{Kim2018} implies a continuous change of carrier concentration near the surface, and therefore a simple two-channel model may not be sufficient, calling for a sophisticated model based on realistic surface states.

The Hall effect is not affected by the thin surface layer and is dominated by the bulk carriers.
The temperature-dependent Hall effect can shed light on the detailed structure of electronic bands near the chemical potential, which is still under debate as the local density approximation \cite{Liu2011, Meinert2016} predicts a significant overlapping of the hole-like and electron-like bands (compensated semimetal) whereas quadratically point-touching bands are predicted by other calculations \cite{Lin2010,Chadov2010,Xiao2010}. 
While the result of our Hall effect measurement cannot precisely predict the band structure, it rules out the compensated semimetal since we do not see any electron-like band at low temperatures.

The surface states affect the superconductivity in YPtBi, and exotic surface superconductivity may coexist with the bulk superconductivity \cite{persky2023possible, schwemmer2022chiral}.
Due to the dimensional and potentially unconventional properties, the surface superconductivity may exhibit distinct characteristics in the upper critical fields.
Strong magnetic fields can generally suppress the superconducting states by two well-known mechanisms. First, superconductivity will be destabilized when free energy reduction due to Pauli spin paramagnetism in a finite magnetic field favors over the superconducting condensation energy\cite{Clogston1962}. 
The Pauli paramagnetic limiting field $H_P$ for spin-1/2 spin-singlet pairing is given by $\Delta_0/\sqrt{2}\mu_B$ for a parabolic band where $\Delta_0$ is the magnitude of the superconducting energy gap at zero temperature and $\mu_B$ is the Bohr magneton. 
For a weak-coupling BCS superconductor, $\Delta_0\approx 1.764 k_B T_c$. 
Therefore, $\mu_0 H_P= a T_c$ where $a \approx 1.86$ T/K. In this approximation, the expected $H_P$ in YPtBi is $\mu_0 H_P\approx 1.53$ T. 

Secondly, the Cooper pair can be de-paired when the increased kinetic energy due to the additional orbital motion from the presence of a magnetic field exceeds the pairing potential energy. 
For a weak-coupling BCS superconductor with a spherical Fermi surface, the orbital pair-breaking effect is best described by $H_{c2}(0)\approx 0.727 |H'_{c2}(T_c)|T_c$ \cite{Helfand1966}, which corresponds to $\mu_0H=1.11$ T for YPtBi. 
Here, we used $\mu_0H'_{c2}(T_c) = -1.98$ T/K. 
Generally, $H_P$ is greater than the orbital $H_{c2}(0)$, and therefore $H_p$ is observed only when the orbital pair-breaking mechanism is significantly weakened.

While a large effective mass naturally leads to a higher orbital-limiting $H_{c2}$ value expectation, quasi-2D systems can exhibit exceptionally high $H_{c2}$ when $H$ is parallel to the plane, limiting the orbital motion. 
In this case, the Pauli limiting field will determine the upper critical field. 
However, the Pauli mechanism can be avoided by accommodating the Cooper pairs with a finite momentum \cite{Fulde1964,Larkin1964}. 
Alternatively, Ising superconductivity due to strong spin-orbit coupling can lead to critical fields far exceeding the Pauli limit \cite{Lu2015}.

The presence of strong spin-orbit coupling \cite{Werthamer1966} or anisotropic superconducting gap structure \cite{Kogan2012} would lead to significant magnetic anisotropic energy which results in the sizable angular variation of $H_{c2}$ in a 3D system.
In such cases, the angle variation of $H_{c2}$ would depend on the orientation of applied $H$ with respect to the crystallographic direction. 
In YPtBi, we observed a large anisotropy in $H_{c2}$ which is irrespective of the bulk crystal symmetry. 
When $H$ is parallel to the plane of the plate-like sample, $H_{c2}$ is consistently higher in various samples, which is reminiscent of angular variation of $H_{c2}$ in a quasi-2D system, and the anomalous electronic phase could exist at the boundary of the sample considering the topological nature of the bulk band structure.

The anomalous transport properties of both normal and superconducting states in YPtBi, possibly arising from topological surface states, would suggest an exotic superconducting state is responsible for the observed high upper critical field. 
The twofold symmetry observed in YPtBi is strikingly similar to evidence for nematic superconductivity reported in experiments on the doped topological insulator Bi$_2$Se$_3$ \cite{Yonezawa2017}, twisted bilayer graphene \cite{Cao2021}, and most recently the kagome superconductor CsV$_3$Sb$_5$  \cite{Xiang2021}. While this seems to be a common feature, isolating the surface and bulk contributions will require more sophisticated experiments such as the double Corbino disc technique \cite{Eo2019}.

\section{summary}

We investigated the anisotropy of transport properties in the normal and superconducting states of the topological semimetal YPtBi. 
The non-metallic electrical resistivity of YPtBi, which rises on cooling along with the Hall coefficient, entails an inherent anisotropy in its response to applied magnetic fields that follows a symmetry reduced from that of the four-fold cubic symmetry of the crystallographic structure. 
The longitudinal resistivity exhibits an anomalous two-fold rotational symmetry up rotation of an applied field perpendicular to the applied electrical current at all temperatures but becomes more pronounced at low temperatures.
In the superconducting state, this reduced symmetry is also apparent in the measured anisotropy of the upper critical field, which also does not follow the rotational symmetry of crystal structure when a magnetic field is rotated out of the sample plane. 
Our results imply the anomalous rotational symmetry-breaking apparent in both the normal and superconducting states of YPtBi cannot be explained by bulk carrier contributions, but must arise from contributions of topological surface states, laying a cornerstone for the realization of time-reversal symmetry-breaking topological surface superconductivity.

\section{acknowledgement}
This work was supported by the U.S. Department of Energy (DOE) Award No. DE-SC-0019154 (experimental investigations), and the Gordon and Betty Moore Foundation’s EPiQS Initiative through Grant No. GBMF4419 (materials synthesis). Undergraduate research (D.K.) was generously supported by the Maryland Quantum Materials Center.

\end{document}